\documentclass[letterpaper,12pt]{article}

\usepackage{anysize}
\marginsize{2.0cm}{2.0cm}{2.0cm}{-0.25cm}

\usepackage{color}
\usepackage{afterpage}
\usepackage{amsmath,amssymb}

\usepackage[numbers,sort&compress]{natbib}
\usepackage{multicol} 

\usepackage{graphicx} % Includes graphics
\usepackage{float} % Allows forcing of figure locations
\usepackage{wrapfig} % Wraps text round images
\usepackage[font=small,format=plain,labelfont=bf,up,textfont=it,up]{caption}

\usepackage{mdwlist} % less gaps in itemize sections
\usepackage{titlesec} % format of section titles
\titlespacing{\section}{0pt}{6pt}{2pt}
\titlespacing{\subsection}{0pt}{6pt}{2pt}
\titleformat*{\section}{\centering \bf \Large}
\titleformat*{\subsection}{\centering \bf \large}
\newcommand{\eg}{e.\,g.}

\usepackage{changepage}   % for the adjustwidth environment

\begin{document}
%%%%%%%%%%%%%%%%%%%%%%%%%%%%%%%%%%%%%%%%%%%%%%%%%
\begin{center}
	{\Large {\bf A hunt for dual radio active galactic nuclei in the VLASS}}\\
\vspace{4mm}
{\normalsize S. Burke-Spolaor$^{1}$\footnote{The North American Nanohertz Observatory for Gravitational Waves (http://www.nanograv.org) notes its support of this proposal. The science described herein would contribute directly to NANOGrav gravitational-wave detection efforts.},~ A. Brazier$^{2}$,~ S. Chatterjee$^{2}$,~ J.~Comerford$^{3}$,~ J. Cordes$^{2}$, J.~Lazio$^4$,~ X.~Liu$^{5}$,~ Y.~Shen$^{6}$\\

\vspace{2mm}
\footnotesize
  $^1$\emph{California Institute of Technology, 1200 E. California Blvd, Pasadena, CA 91125}\\
  $^2$\emph{Department of Astronomy, Cornell University, Ithaca, NY 14853}\\
  $^3$\emph{Department of Astrophysical and Planetary Sciences, University of Colorado, Boulder, CO 80309}\\
  $^4$\emph{Jet Propulsion Laboratory, California Institute of Technology, 4800 Oak Grove Dr., Pasadena, CA 91109}\\
  $^5$\emph{Division of Physics \& Astronomy, UCLA, Los Angeles, CA 90095}\\
  $^6$\emph{Carnegie Observatories, 813 Santa Barbara Street, Pasadena, CA 91101}\\
}

\end{center}
\vspace{-5mm}
\section*{Abstract}
\begin{quote}
This whitepaper describes how the VLASS could be designed in a manner to allow the identification of candidate dual active galactic nuclei (AGN) at separations $<$7\,kpc. Dual AGN represent a clear marker of two supermassive black holes within an ongoing merger.  A dual AGN survey will provide a wealth of studies in structure growth and gravitational-wave science.
Radio wavelengths are ideal for identifying close pairs, as disturbed stellar and gaseous material can obscure their presence in optical and shorter wavelengths.
% This is not true with X-rays of course but whatever; Chandra's not about to do a deep all-sky survey. 
With sufficiently high resolution and sensitivity, a large-scale radio imaging survey like the VLASS will uncover many of these systems and provide the means to broadly study the radio properties of candidate dual systems revealed at other wavelengths.  
We determine that the ideal survey for our purposes will be at as high a resolution as possible, with significantly more science return in A array at L-band or higher, or B array at C-band or higher.
We describe a range of potential survey parameters within this document. Based on the analysis outlined in this whitepaper, our ideal survey would create a catalogue of $\gtrsim$100 dual AGN in either: 1) a medium-sensitivity ($\sim$1\,mJy detection threshold), wide-field (few thousand square degree) survey, or 2) a high-sensitivity ($\sim$10\,$\mu$Jy threshold) survey of several hundred square degrees.
%A large-sky radio survey for dual SMBHs will provide a wealth of science, exploring: 
% - AGN ignition relationship to merger events.
% - SMBH growth via accretion during merger.
% - Predicted significance of gas influence on GW-emitting binaries.
% - Multi-messenger science possibilities
% - Galaxy merger statistics.
\end{quote}

%Radio source counts at uJy
%http://adsabs.harvard.edu/abs/2012ApJ...758...23C
%NVSS derived source counts:
%http://adsabs.harvard.edu/abs/2013ApJ...768...37C

\vspace{-5mm}
\section{Background}\label{sec:intro}
\noindent Dual ($\lesssim$ 10\,kpc separation) and binary ($\lesssim$ 100\,pc separation) supermassive black holes (SMBHs) are formed during the merger of two massive galaxies. Hierarchical structure formation scenarios imply that $\sim$0.1--1\% of galaxies might harbor a dual SMBH at redshifts $z<2$ \citep[\eg][]{VHM}. If gas is available in a merger, accretion onto the SMBHs may occur, powering a dual active galactic nucleus that can be identified via its radio emission. 

The frequency of dual SMBHs in galaxy mergers is central to our understanding of galaxy evolution, and to our predictions for the strength of gravitational-wave signals for pulsar timing arrays and future space-based laser interferometers. Determining the rate of dual \emph{active nuclei} in galaxy mergers will explore merger-induced active nucleus activity and SMBH growth.

Dual SMBHs discovered at separations within typical galaxy virialization radii ($\sim$7\,kpc) represent confident detections of imminent mergers, therefore represent a strong probe of redshift-dependent merger rates, occupation fractions of dual SMBHs in galaxies, and post-merger dynamical evolution. These measurements feed directly into gravitational wave signal predictions for pulsar timing arrays.
Within this radius, dual AGN detections probe merger-induced SMBH growth, and a large fraction of dual AGN are anticipated due to gas infall during the merger \citep[\eg][]{koss12}.
Studies of dual nucleus systems can identify features unique to recent galaxy mergers that
provide easy markers of such systems; this may enable direct confirmation of gravitational-wave discoveries with pulsar timing arrays \cite{sbs-cqg}.

Despite the broad interest in SMBH pair identification, few SMBH pairs within a common galactic envelope are known. The number of known systems decreases at smaller separations, primarily due to difficulties in their identification at short wavelengths. The VLA provides an efficient and high-resolution means to discover dual SMBHs, however previous surveys (NVSS, FIRST, etc.) did not reach sufficient sensitivity or resolution for a dual nucleus search.

In this whitepaper, we detail how the upcoming VLASS may be used as a key tool for dual SMBH science if our proposed resolution and sensitivity requirements can be met.

\begin{figure*}
\centering
\vspace{-2mm}
\includegraphics[width=0.43\textwidth,trim=0mm 7mm 0mm 5mm,clip]{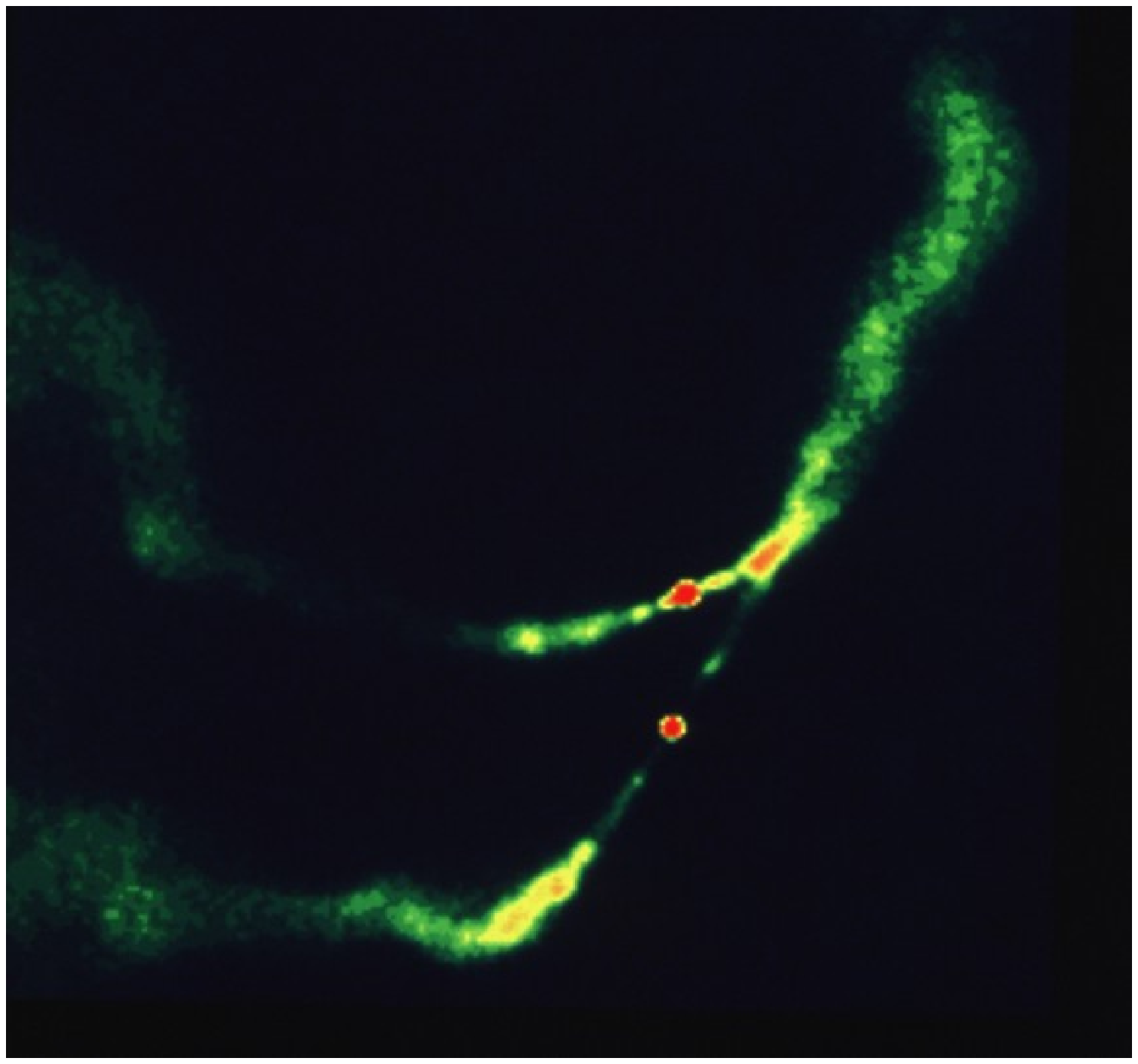}~~~~~
\includegraphics[width=0.43\textwidth,trim=0mm 15mm 0mm 10mm,clip]{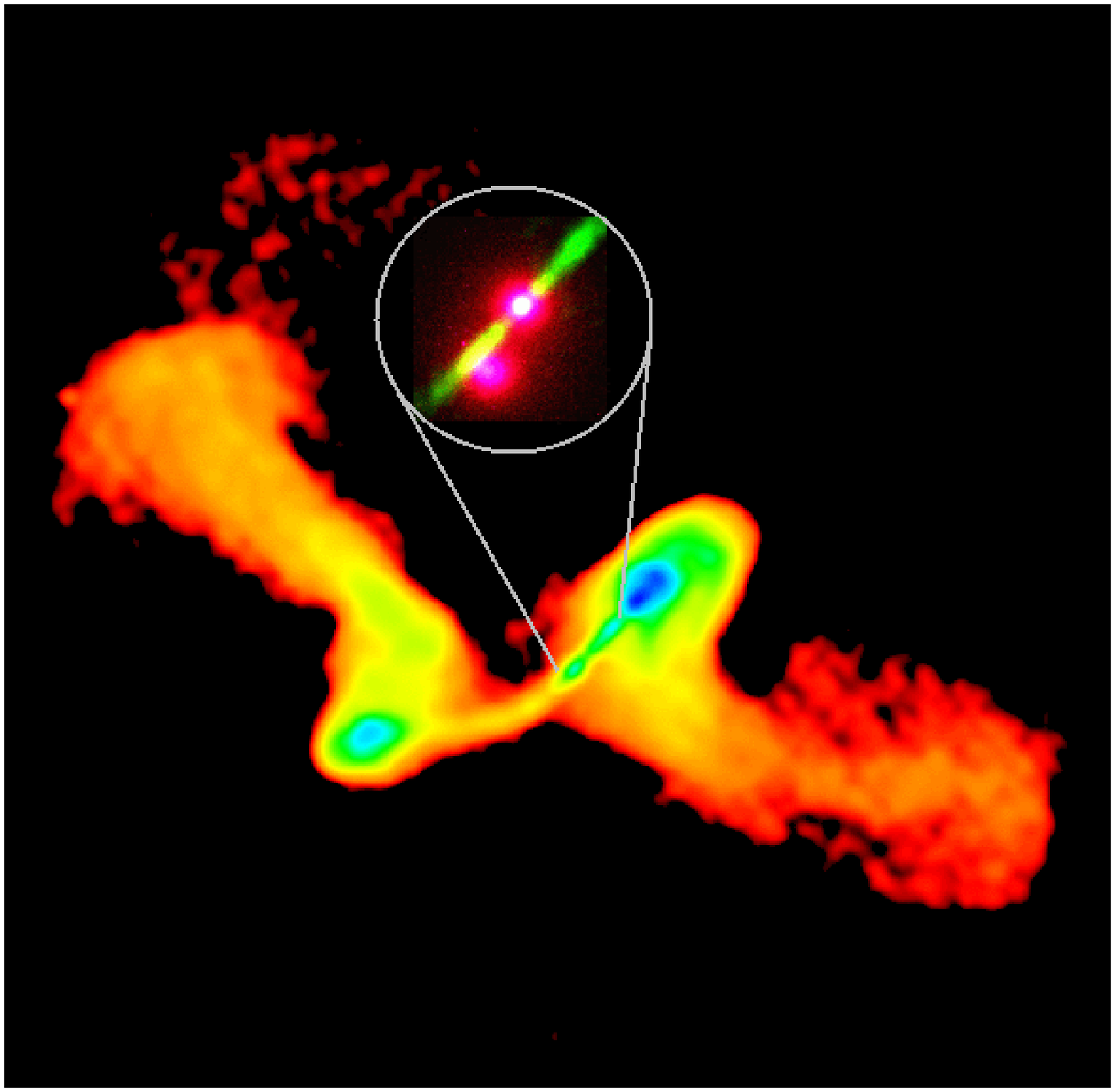}
\vspace{-2mm}
\caption{Two dramatic examples of dual active galactic nuclei identified through their radio emission. \textbf{Left:} 3C75, whose dual cores have a 7\,kpc projected separation, exhibits large-scale jet structures that have misaligned but correlated morphology. Image courtesy of NRAO/AUI and F.N. Owen, C.P. O'Dea, M. Inoue, \& J. Eilek. \textbf{Right:} NGC326, which has an X-shaped structure that inspired its dual nucleus interpretation (which was confirmed at other wavelengths, as shown). Image courtesy of NRAO/AUI, STScI (inset).
}\vspace{-3mm}
\label{fig:3c75}
\end{figure*}

\section{The Search for Dual Radio SMBHs}\label{sec:search}

\subsection{Identification techniques}\noindent 
To find dual SMBH candidates, we must analyze the morphology, distribution, and spectra of closely-separated radio components.
As individual nuclei might be core or jet/lobe-dominated, to identify a dual AGN we would look for any of the following features:
\begin{itemize}
\vspace{-1mm}\item Two flat-spectrum cores within a proximity of a few arcseconds.
\vspace{-2mm}\item Multiple apparent jets with misaligned axes (\eg\ X-shaped objects; Fig.\,\ref{fig:3c75}).
\vspace{-2mm}\item A flat-spectrum object nearby, but not aligned with the axis of, an extended source.
%\vspace{-1.5mm}\item ...???
\vspace{-1mm}
\end{itemize}
\noindent While such objects can be automatically identified, manual inspection will be required to determine contaminating sources (\eg\ wide-angle tail galaxies). This project would lend itself excellently to a Galaxy Zoo-style project, which would form an excellent public outreach platform for radio astronomy. We have reached out to Galaxy Zoo, and they have agreed that this project represents an excellent example of citizen science, if the angular resolution of the VLASS is sufficiently high to produce a suitable number of candidates.

Discoveries would be followed up (\eg\ with the VLBA or higher-resolution VLA) and cross-matched with optical surveys to measure spectra, redshifts, core separations, higher-resolution radio morphology, and disturbed host galaxy structures to determine whether the pair is a genuine dual AGN, a gravitational lens, or a chance projection. Similar studies have been performed for all successful dual AGN detections, as in Fig.\,\ref{fig:3c75} and \cite{rodriguez06}.

\subsection{Scales of interest}\label{sec:scales}
\begin{figure}
\centering
\vspace{-5mm}
\includegraphics[angle=270,width=0.99\textwidth,trim=0mm 0mm 0mm 0mm,clip]{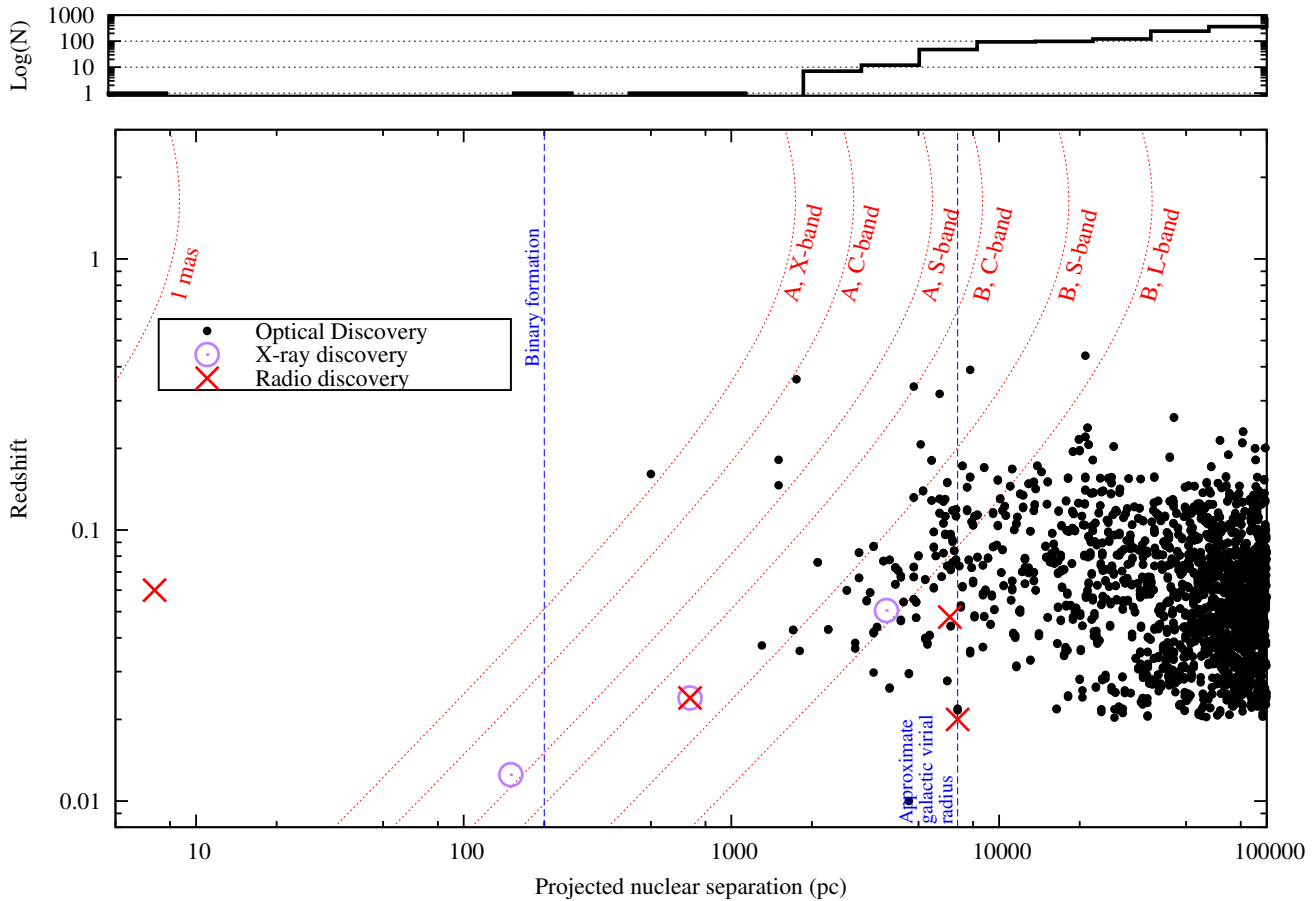}
\vspace{-2mm}
\caption{Each point shown represents a dual AGN. Dots, red X's, and purple circles mark optical \cite{liu11}, radio, and X-ray discoveries, respectively. A typical galactic virial radius, and the separation at which two SMBHs are thought to pair, are indicated. The upper panel shows the integrated number of discoveries as a function of nuclear separation; optical identification becomes far less effective at separations below the galactic virial radius. The dotted curves mark the resolution limit of dual AGN discovery for several VLA survey configurations. A-array at L-band is close to the B-array C-band curve; these configurations and those at higher resolution will have the highest science return, as they open up significant discovery space below the galactic virial radius.
The 1\,mas curve marks the large discovery space accessed by the VLBA, as it can perform follow-up of large-scale VLA structures identified as dual AGN indicators (\eg\ X-shaped jets or periodic morphology).}
\label{fig:vlares}
\end{figure}

\noindent
The VLA has unmatched potential for dual SMBH discovery in is its combination of resolution and survey speed. 
%Its nearest competitor is perhaps Chandra X-ray Observatory, at $\sim$0.5$''$, which has discovered a number of dual nuclei in targeted pointings.
We aim to harness these strengths to probe dual nuclei at separations difficult to access with optical telescopes. Based on past discoveries, summarized in Fig.\,\ref{fig:vlares}, optical detection rates begin to falter around 10\,kpc, worsening significantly below a massive galaxy virial radius of $\sim$7\,kpc. The VLA has huge discovery space within this radius.
Depending on SMBH masses and their hosts' environments, SMBHs can form a bound binary once they are within around 1--200\,pc separations. Systems within this radius are \emph{direct precursors} to the gravitational-wave emitters detectable by pulsar timing arrays and space-based laser interferometers, so are of utmost interest.

\begin{figure}
\centering
\vspace{-3mm}
\includegraphics[width=0.6\textwidth,trim=20mm 20mm 12mm 21mm,clip]{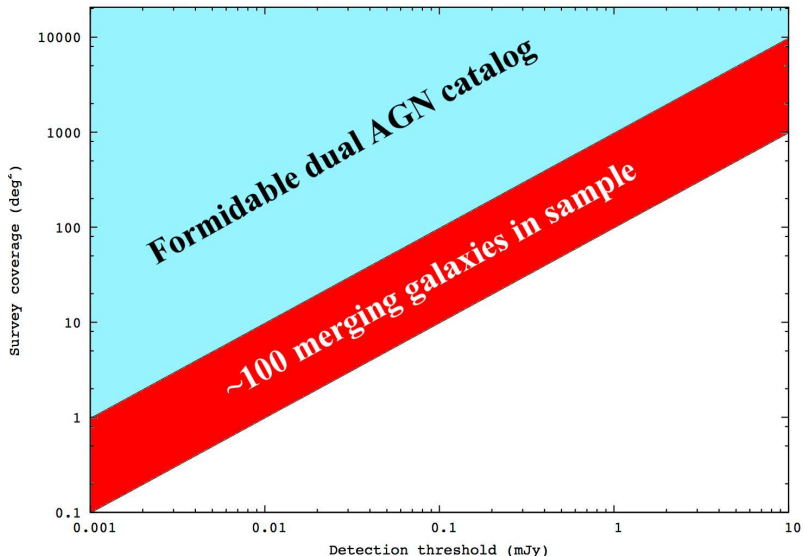}
\vspace{-3.5mm}
\caption{Here we show the trade-off between sky sensitivity and survey coverage for discovering dual AGN. The construction of the red curve is described in \S\ref{sec:sens}. Pairs should spend the most time at the large scales probed by the VLA, however the actual discovery rate in this space will depend on how effectively mergers induce radio AGN activity, and SMBH evolution efficiency. We therefore consider any survey region above this curve sufficient to construct a significant dual AGN catalog.}
\label{fig:sky}
\end{figure}

\subsection{Ideal array configuration and frequency}\noindent
\noindent Resolution is the primary science delimiter for our search: better resolution will lead to higher source counts and greater discovery impacts (\S\ref{sec:intro},\ref{sec:scales}). Figure~\ref{fig:vlares} summarizes the discovery space probed by a number of survey configurations.
A-array at L-band or higher frequency, or B-array at C-band or higher, will probe within the virial radius at nearly all redshifts.

\subsection{Anticipated discovery rates vs.\ sensitivity and sky coverage}\label{sec:sens}\noindent
A dual AGN catalog of at least one hundred conclusive detections would form a significant science database.
We estimate our ideal sensitivity and sky coverage as follows. For a survey of discovery threshold $S$, we take a representative number of radio sources per square degree as given by the integral radio source counts of \cite{windhorst} at 1.4\,GHz for $S>275\,\mu$Jy, and extrapolate these using a scaling of $S^{-1}$ to lower flux densities. The number of \emph{ongoing mergers} in a blind discovery sample
will be $\sim$0.1--1\% of discoveries. Therefore, we scale the source count by this factor to estimate the number of potential dual AGN per deg$^2$. For a maximum of 100 dual AGN discoveries, we show the required sky coverage and sensitivity in Figure\,\ref{fig:sky}. Because not all dual AGN may be radio-emitting, our survey discovery impacts will have the most return above this threshold.

\section{Example studies and impacts}\noindent
Here is an example list of studies that our proposed search could be used for; there are more, however these represent the current interests of the authors of this whitepaper:
\begin{itemize}
\vspace{-1.5mm}\item {\bf Measure merger-induced SMBH growth via accretion}.
It is known that galaxy mergers can fuel black hole mass growth, but it is still unclear when mass growth peaks in a merger.  Simulations show mergers driving gas to the centers of merger-remnant galaxies \citep[\eg][]{springel05}, so there is an expectation that the bulk of a black hole's mass growth occurs when it nears the center of the merger-remnant. Simulations predict that the instantaneous black hole growth rate peaks around pair separations of 1--10\,kpc \citep{VanWassenhove12} or 0.1--2\,kpc \citep{Blecha13}. Observations have indicated an increasing active nucleus fraction down to 10\,kpc \citep{koss12}, but the trend is not yet clear at smaller separations because of the dearth of known nuclei at $<$10 kpc separations. 
\vspace{-1.5mm}\item {\bf Predict the strength of the gravitational-wave background, continuous waves, and burst signals for pulsar timing arrays}. Flux-limited radio active nucleus surveys tend to find the most massive SMBHs, which are the SMBHs that are most important in predictions of the gravitational wave signals detected by pulsar timing arrays. The VLASS's dual nucleus census would provide a strong assessment of the rates and detectability of SMBH binaries for pulsar timing arrays. This is a particularly timely measurement, as the limits from timing arrays are beginning to breach predictions from standard hierarchical cosmology predictions for SMBH growth \citep{pptascience}, and the first expected gravitational-wave detection may come as soon as 2016 \citep[\eg][]{siemens-CQG}.
\vspace{-1.5mm}\item {\bf Identify large-scale multi-messenger indicators of binary SMBHs}. Performing multi-wavelength (new and archival) studies of dual active nucleus systems will help identify emission across the spectrum that are unique to galaxies containing two SMBHs at late stages of merger evolution. This will be an invaluable tool for future multi-messenger science with gravitational-wave detections \citep[\eg][]{SathyaprakashSchutz09,sbs-cqg}. As with the previous point, this project will be particularly timely in the coming 5--10\,years.
\vspace{-1.5mm}\item {\bf Cross-match the VLA's dual active nucleus catalog with optical dual nucleus discoveries.}
As seen in Fig.\,\ref{fig:vlares}, many optical surveys have sought dual nuclei through emission-line diagnostics. A VLASS radio active nucleus search will provide a vast database for use as a control sample, and for multi-wavelength assessment, complementing the foundational work of \eg\ \cite{comerford09,liu2010,shen13} in this field.
To a certain extent, the NVSS and other previous large-scale radio surveys suffered from resolution and sensitivity issues for such comparisons. For instance, a cross-correlation of the FIRST survey and a large optical dual AGN catalogue \cite{liu11} finds only $\sim$100 matches. Thus, to be effective the VLASS would have to be at least deeper than the FIRST survey, and wide sky coverage would enable more significant cross-matching with optical catalogs.
If an off-galactic-plane component to the VLASS has significant overlap with the SDSS, the improved VLASS resolution will enable confident association of radio sources with a galactic host.
\vspace{-1.5mm}\item {\bf Radio and multi-wavelength follow-up}. Systematic follow-up of X-shaped or peculiar-morphology dual AGN candidates in the VLASS database could be done with the VLBA to reveal the presence of dual cores below the resolution limit of the VLA. This could potentially reveal bound binary SMBH systems. Other wavelength measurements could pursue additional science goals: 
ALMA observations could compare with a control (single-AGN) sample to identify any differences in the gas environment of the systems, and optical morphology studies of close dual AGN could indicate how advanced the host merger appears to be.
\end{itemize}

\vspace{-2mm}
% The reference list...
\bibliographystyle{unsrt}
\def\bibfont{\scriptsize}
\begin{multicols}{2}
\bibliography{vlass-dual-agn}
\end{multicols}

%\subsection*{Acknowledgements}\textit{}
\end{document}